\documentstyle[preprint,prl,aps,]{revtex}
\tightenlines

\begin{document}
\draft
\title{Electron and orbital correlations in Ca$_{2-x}$Sr$_{x}$RuO$_{4}$ probed by
optical spectroscopy}
\author{J. S. Lee,$^1$ Y. S. Lee,$^2$ T. W. Noh,$^{1,}$\cite{email} S.-J. Oh,$^2$
Jaejun Yu,$^2$ S. Nakatsuji,$^{3,}$\cite{address} H. Fukazawa,$^3$ and Y.
Maeno$^3$}
\address{$^1$School of Physics and Research Center for Oxide Electronics,\\
Seoul National University, Seoul 151-747, Korea}
\address{$^2$ School of Physics and Center for Strongly Correlated\\
Materials Research, Seoul National University, Seoul 151-747, Korea}
\address{$^{3}$Department of Physics, Kyoto University, Kyoto 606-8502, Japan}
\date{July 17, 2002}
\maketitle

\begin{abstract}
The doping and temperature dependent optical conductivity spectra of the
quasi-two-dimensional Ca$_{2-x}$Sr$_x$RuO$_4$ (0.0$\leq ${\it x}$\leq $2.0)
system were investigated. In the Mott insulating state, two electron
correlation-induced peaks were observed around 1.0 and 1.9 eV, which could
be understood in terms of the 3-orbital Hubbard model. The low frequency
peak showed a shift toward higher frequency as temperature was lowered,
which indicated that electron-phonon interactions play an important role in
the orbital arrangements. From the systematic analysis, it was suggested
that the antiferro-orbital and the ferro-orbital ordering states could
coexist.
\end{abstract}


\newpage
The discovery of unconventional superconductivity in Sr$_2$RuO$_4$ has
stimulated great interest in layered perovskite Ca$_{2-x}$Sr$_x$RuO$_4$ (0.0$%
\leq ${\it x}$\leq $2.0) (CSRO) \cite{Maeno nature,Nakatsuji PRL}. The
normal state of Sr$_2$RuO$_4$ can be described as a quasi-two dimensional
Fermi-liquid \cite{Fermi liquid}; however, the isovalent Ca substitution for
Sr gives rise to a Mott insulator Ca$_2$RuO$_4$ \cite{Ca214 Mott}. As {\it x}
decreases, the RuO$_6$ octahedra rotate and become flattened \cite{Friedt}.
These structural distortions decrease the Ru 4$d$ bandwidth $W$, which leads
to a Mott gap. Therefore, CSRO can be used as a prototype material system to
investigate the evolution of electronic structures from a band metal to a
Mott insulator.

The role of the orbital degrees of freedom in Ca$_2$RuO$_4$ has provided us
with a challenge of scientific interest. The Ru ion has a formal valency of
4+ with four $t_{2g}$ electrons. At an early stage, it was proposed that the 
$d_{xy}$ orbitals should be fully occupied and that a Mott gap becomes open
between the half-filled $d_{yz/zx}$ states \cite{Anisimov}. However, recent 
{\it x}-ray absorption spectroscopy (XAS) data revealed that both $d_{xy}$
and $d_{yz/zx}$ orbitals remain partially filled down to low temperature ($T$%
) \cite{Mizokawa}. Then, it would be difficult to understand how the Mott
insulator can be formed with partially filled $d_{xy}$, $d_{yz}$, and $%
d_{zx} $ orbitals. More recently, using the 3-orbital Hubbard model, Hotta
and Dagotto suggested that the unusual orbital occupancy could be explained
by an antiferro-orbital (AFO) ordering in the antiferromagnetic (AFM) state 
\cite{OO}.

Optical spectroscopy has been used as a powerful method to investigate
electronic structures of numerous strongly correlated materials \cite{Cooper}%
. However, most of the observed correlation-induced features have been
interpreted in terms of the single-band Hubbard model without considering
the orbital degeneracy. Recently, optical spectroscopy was applied to probe
the role of orbital ordering in polaron absorption of perovskite manganites;
even with similar orbital occupancy, optical absorption can vary
significantly depending on the orbital correlation between nearest neighbors 
\cite{MWKim}. Therefore, optical spectroscopy can provide us with a way to
investigate the role of orbitals in the intriguing electronic states of CSRO.

In this Letter, we investigate the doping- and $T$-dependent optical
conductivity spectra $\sigma (\omega )$ of CSRO. In the insulating state, $%
\sigma (\omega )$ show two correlation-induced excitations around 1.0 and
1.9 eV. These features are explained in terms of the 3-orbital Hubbard
model, which considers the $t_{2g}$ orbital multiplicity. In the Mott
insulating state, the AFO correlation is prevalent, but the ferro-orbital
(FO) correlation seems to coexist with decrease of $T$. It is also found
that the electron-phonon interaction plays an important role in the $T$%
-dependent evolution of the orbital arrangements.

High-quality CSRO single crystals were grown by the floating zone method.
Details of the crystal growth and characterization were described elsewhere 
\cite{Ru214 growth}. Near normal incident reflectivity spectra, $R(\omega )$%
, of the $ab$-plane were measured between 5 meV and 30 eV. Using a liquid
He-cooled cryostat, $T$-dependent $R(\omega )$ were obtained below 5.0 eV.
Using Kramers-Kronig analysis, $\sigma (\omega )$ were obtained \cite{KK}.

Figure 1 shows $\sigma (\omega )$ of CSRO at room $T$. For Sr$_2$RuO$_4$, $%
\sigma (\omega )$ has a strong Drude-like peak, indicating its metallic
state. As {\it x} decreases, the Drude-like peak becomes gradually
suppressed. For Ca$_2$RuO$_4$, $\sigma (\omega )$ shows an insulating
response with an optical gap. These spectral changes are consistent with the
metal-insulator (MI) transition observed in {\it dc} resistivity $\rho _{dc}$
\cite{Nakatsuji PRL}. The strong peak around 3.0 eV can be assigned to the $%
p $-$d$ transition from the O 2$p$ to the unoccupied Ru $t_{2g}$ states,
just like in earlier works on numerous ruthenates \cite{ruthenates
sig,Puchkov,Tokura214}. Therefore, the spectral features below 2.5 eV should
originate from transitions between the Ru 4$d$ bands.

The spectral weight changes in Fig. 1 can be understood according to the
Mott-Hubbard picture. As {\it x} decreases, the structural distortions
become larger \cite{Friedt}, which makes $W$ of the Ru 4$d$ orbitals
narrower. As $U/W$ increases ($U$ is the electron correlation energy.), the
Hubbard bands develop with reduction of the quasi-particle peak centered at
the Fermi energy \cite{RMP 1998}. Figure 1 shows that an additional peak
develops around 2.0 eV as {\it x} decreases. To identify this peak more
easily, each spectrum is subtracted by $\sigma (\omega )$ of {\it x}=2.0 and
plotted as $\Delta \sigma (\omega )$ in the inset of Fig. 1. The spectral
weight redistributions between the coherent and the incoherent excitations
can be understood in the $W$-controlled Mott-Hubbard picture. Then, the 2.0
eV excitation should be assigned as the so-called $`U$-peak', i.e.{\it ,}
the optical transition from the lower Hubbard band to the upper Hubbard band.

The increase of $U/W$ should lead to an AFM Mott insulator. For its $\sigma
(\omega )$, only one correlation-induced peak\ was usually reported \cite
{RMP 1998}. However, $\sigma (\omega )$ of the {\it x}=0.0 sample exhibits
two separate peaks around 1.0 and 1.9 eV. Earlier, Puchkov {\it et al.}
reported $\sigma (\omega )$ of Ca$_2$RuO$_4$ up to 1.3 eV and assigned the
low energy peak to the $U$ peak \cite{Puchkov}. The high energy peak
corresponds to the 2.0 eV excitation in the metallic CSRO. If both peaks
come from the electron correlation effects, the single-band Hubbard model
cannot explain the appearance of these two peaks. Moreover, the simple
picture cannot explain the recent XAS studies, which reported a dramatic
redistribution of orbital occupancies \cite{Mizokawa}.

To obtain further insights, we investigated $\sigma (\omega )$ of the {\it x}%
=0.06 sample, which undergoes an MI transition below room $T$ \cite{Comment
exp1}. The inset of Fig. 2(a) shows that $\rho _{dc}$ experiences an abrupt
jump at $T_{MI}\sim $ 220 K with decreasing $T$, which is closely related to
the first-order structural transition \cite{Friedt}. Figure 2(a) shows the $%
T $-dependent $\sigma (\omega )$, where the sharp spikes below 0.1 eV are
due to transverse optic phonons. Consistent with $\rho _{dc}$, $\sigma
(\omega )$ also show the first-order nature of the MI transition; below $%
T_{MI}$, the strong coherent peak suddenly disappears and an optical gap
appears with the 1.0 eV peak (Peak $\alpha $), accompanied by a slight
enhancement of the 1.9 eV peak (Peak $\beta $). The new Peak $\alpha $ does
exist in the insulating region of the {\it x}=0.06 sample as well as in the
room-$T$ $\sigma (\omega )$ of the {\it x}=0.0 sample. [Note that $T_{MI}$
for the {\it x}=0.0 sample is around 357 K \cite{Cao Ca214}.] Thus, the
two-peak structure can be regarded as an ubiquitous feature of the Mott
insulating state in CSRO.

Note that there exist systematic spectral changes below 2.5 eV even in the
insulating state. In order to illustrate these changes clearly, we
subtracted the contributions of the phonons and the $p$-$d$ transitions,
indicated as a dot-dot-dashed line in Fig. 2(a), from each spectrum to
obtain the contribution of the transitions between 4$d$ states, $\sigma
_{4d}(\omega )$. Figure 2(b) shows the two-peak structure of $\sigma
_{4d}(\omega )$. As $T$ decreases, the spectral weight of Peak $\alpha $ ($%
S_\alpha $) decreases, but that of Peak $\beta $ ($S_\beta $) increases. The
sum of the spectral weights up to 3.0 eV remains the same, satisfying the
optical sum rule. It should be noted that only Peak $\alpha $ shows the $T$%
-dependent peak shift.

To explain these spectral changes, we consider the 3 $t_{2g}$-orbital
Hubbard model. When the orbital multiplicity is taken into account, the
Hund's rule exchange energy $J$ \ between $t_{2g}$ electrons should play an
important role. The interactions among electrons on the same site can be
described by $U$ Coulomb repulsion energy (CRE) between two electrons in the
same orbital, $U^{\prime }$ CRE between two electrons in different orbitals
with opposite spins, and $U^{\prime \prime }$ CRE between electrons in
different orbitals with the same spin. [When the rotational symmetry in the
orbital space is considered, $U^{\prime }$= $U-2J$, and $U^{\prime \prime }$%
= $U-3J$ \cite{Degenerate HM}.] As shown in Fig. 3, the energy cost of the $%
d $-$d$ transition (i.e., $d^4$ + $d^4\longrightarrow $ $d^3$ + $d^5$)
depends on the spin and the orbital configurations of the nearest
neighboring Ru ions. For the time being, let us assume that all of the $%
t_{2g}$ states are degenerate. Then, we should consider four kinds of
spin/orbital configurations. Figure 3(a), (b), (c), and (d) show possible $d$%
-$d$ transitions in FM (ferromagnetic)/FO, FM/AFO, AFM/FO, and AFM/AFO
configurations, respectively. [Here, FO (or AFO) implies that the same (or
different) $t_{2g}$ orbitals are occupied in the two sites.] With the above
mentioned CRE's, the corresponding energy costs of the $d$-$d$ transitions
are estimated. As shown in Fig. 3, the FM/FO configuration does not allow
any $d$-$d$ transition. And, the FM/AFO and the AFM/FO configurations allow
only one transition. On the other hand, the AFM/AFO configuration allows two
transitions with energy costs of $U-J$ and $U+J$ \cite{Comment1}.

How can we understand the two-peak structure? Note that the {\it N\'{e}el}
temperature $T_N$ of the {\it x}=0.06 sample is around 150 K \cite{Nakatsuji
PRB}. Near and below $T_N$, the nearest neighbor spin correlation is mostly
AFM, so most excitations come from the AFM/AFO and the AFM/FO
configurations. Then, Peaks $\alpha $ and $\beta $ can be assigned as
excitations of $U-J$ and $U+J$, respectively. From the energy difference
between two excitations, the value of $2J$ can be estimated to be about 1
eV. It is in good agreement with the earlier reports \cite{J in ruthenates},
thus demonstrating the validity of our peak assignments.

The $T$-dependent spectral weight changes in Fig. 2(b) can be understood as
a result of changes of the AFO and the FO correlations. Using the Lorentzian
fittings, we estimate $S_\alpha $ and $S_\beta $ \cite{Comment Fit}. Since
the value of $S_\alpha $ is similar to that of $S_\beta $ at most
temperatures, the contribution of the AFO correlation seems to be more
prevalent than that of the FO correlation. Figure 4 shows the $T$-dependent
spectral weight changes of $S_\alpha $ and $S_\beta $. As $T$ decreases, $%
S_\alpha $ decreases, but $S_\beta $ increases. These $T$-dependences
suggest that the FO correlation increases with decreasing $T$. Then, it can
be said that there is a competition between AFO and FO configurations at
least in the ground state of the {\it x}=0.06 sample.

Another interesting feature is that Peak $\alpha $ shifts to higher
frequencies by about 0.24 eV with decreasing $T$, but Peak $\beta $ does not
shift significantly. Note that $U$ and $J$ values for most solids have
little $T$-dependences. To explain this intriguing behavior, we pay
attention to the electron-phonon ($el$-$ph$) coupling. As $T$ decreases, it
was known that the RuO$_6$ octahedra rotate and become flattened \cite
{Friedt}. The inset of Fig. 2(b) shows the phonon spectra of the in-plane
Ru-O stretching mode \cite{Phonon}. As $T$ decreases, the phonon frequency $%
\omega _{TO}$ decreases rather significantly, which suggests an increase of
the in-plane Ru-O bond length (i.e., the flattening of the octahedron). In
addition, the strong $el$-$ph$\thinspace coupling makes the phonon lineshape
asymmetric, in agreement with Raman measurements \cite{el-ph}. Figure 4 also
shows that the changes of $\omega _{TO}$ and $\Delta S_\alpha (T)$ could be
related.

Now, let us consider the $el$-$ph$ coupling energy $E_{ph}$. Note that the
Ru ions with 4 $t_{2g}$ electrons cannot have degenerate $t_{2g}$ levels.
Due to the strong Jahn-Teller-like distortion, the energy levels of the
doubly and the singly occupied orbitals will change by -2$E_{ph}$ and $%
E_{ph} $, respectively \cite{OO}. For example, let us consider the orbital
polarized state, shown in Fig. 3(e), which is composed of two sites with
doubly occupied $d_{xy}$ and $d_{yz}$ orbitals. Its schematic energy level
with the Jahn-Teller-like distortion is shown in Fig. 3(f). Compared to the
degenerate case, the orbital polarized state with the AFM/AFO configuration
will have an additional energy cost of 3$E_{ph}$ in the $U-J$ excitation, so
the position of Peak $\alpha $ should be located at $U-J+3E_{ph}$. However,
the $U+J$ excitation contributing to Peak $\beta $ remains independent of
the lattice distortion. With $E_{ph}\sim \omega _{TO}$ $\sim $ 600 cm$^{-1}$
($\simeq $ 0.074 eV), the shift of Peak $\alpha $ is estimated to be about
0.22 eV, in agreement with our experiment. The continuous shift of the $U-J$
excitation\ can be explained by the fact that short range orbital orderings,
which accompanies the lattice distortion in Fig. 3(e), grows progressively.
Near $T_N$, corresponding long range orbital orderings become stabilized and
most of the spectral weights will move to $U-J+3E_{ph}$. Our work
demonstrates that the $el$-$ph$ coupling should be strong in CSRO and play
an important role in the evolution of the orbital degree of freedom. The
shift of $\omega _{TO}$, due to the stong $el$-$ph$ coupling, should be
closely related with the orthorhomic distortion. This distortion will
suppress the AFO ordering \cite{Anisimov} and can explain the close relation
between $\omega _{TO}$ and $\Delta S_\alpha (T)$, observed in Fig. 4.

To explain the electronic structure of the Ca$_2$RuO$_4$ Mott insulating
state, Anisimov {\it et al.} proposed an FO ordered state of the $d_{xy}$
orbitals \cite{Anisimov}. This intriguing state can explain a Mott gap in
the half-filled $d_{yz/zx}$ state and an orthorhombic lattice distortion.
Recently, Hotta and Dagotto proposed another interesting AFO ordered state,
stabilized by a combination of the correlation and the $el$-$ph$ coupling
effects \cite{OO}. In the AFO ordered state (shown in Fig. 1(e) of Ref. \cite
{OO}), the two-dimensional network composed of $d_{xy}$ orbitals is occupied
in a bipartite manner by half of the excess $t_{2g}$ electrons. The other
excess electrons occupy $d_{yz}$ and $d_{zx}$ orbitals. The related lattice
distortions should occur in pairs, resulting in a cooperative distortion in
the total in-plane lattice. Our observation of the prevalence of the AFO
correlation and the energy shift of the $U-J$ excitation seem to be
consistent with the latter picture. However, the increase of the FO at lower 
$T$ cannot be simply explained.

One possible scenario is the phase coexistence of the FO and the AFO ordered
states in the insulating CSRO. In the two-dimensional square network, the
AFO ordered state will be preferred, since it provides the kinetic energy
gain by allowing hoppings between the nearest neighbors \cite{OO}. On the
other hand, with the orthorhombic lattice distortion, the FO ordered state
will be preferred \cite{Anisimov}. If the energy difference between two
states is small, the FO and the AFO ordered Mott insulating states can
coexist. Then, the insulating CSRO will be in an interesting state composed
of two kinds of Mott insulators. Further works are required to verify this
intriguing possibility.

In summary, we investigated the doping- and temperature-dependent optical
conductivity spectra of Ca$_{2-x}$Sr$_x$RuO$_4$. Based on the 3-orbital
Hubbard model, we could explain the two correlation-induced excitations.
Moreover, from their spectral weight changes, we found that the
electron-phonon coupling plays an important role in the dynamics of the
spin/orbital correlations. We also proposed the possible phase coexistence
of orbital ordered states in the insulating samples.

We would like to thank Y. Chung, J.-H. Park, Y. K. Bang, and T. Hotta for
their help and discussions. We also acknowledge the Pohang Advanced Light
Source for allowing us to use some of their facilities. This work was
supported by the Ministry of Science and Technology through the Creative
Research Initiative program and by KOSEF through the Center for Strongly
Correlated Materials Research.

\begin{figure}[tbp]
\caption{Doping-dependent $\sigma (\omega )$ of Ca$_{2-x}$Sr$_x$RuO$_4$ at
room temperature. $\sigma (\omega )$ of {\it x}=0.0 is shown in the lower
panel shifted downward by 500 $\Omega ^{-1}$cm$^{-1}$. The inset shows $%
\Delta \sigma (\omega )=\sigma (\omega )-\sigma (\omega )_{x=2.0}$. }
\label{Fig:1}
\end{figure}
\begin{figure}[tbp]
\caption{(a) $T$-dependent $\sigma (\omega )$ of $x$=0.06. The contribution
of the $p$-$d$ transitions is indicated as a dot-dot-dashed line. The inset
shows $T$-dependent $\rho _{dc}$. (b) The subtracted $\sigma (\omega )$ of
the insulating state. Contributions of the $p$-$d$ transitions and phonons
were subtracted from each spectrum. The inset shows softening of a phonon
peak with decreasing $T$. }
\label{Fig:2}
\end{figure}
\begin{figure}[tbp]
\caption{The spin/orbital configurations of $t_{2g}$ orbitals at the two
lattice sites: (a) FM/FO, (b) FM/AFO, (c) AFM/FO, and (d) AFM/AFO. Possible
optical excitations are also shown in each configuration. (e)
Jahn-Teller-like structural distortion induced by the doubly occupied $%
d_{xy} $ and $d_{zx}$ orbital with the AFO configuration. Octahedron at the
left and the right side are flattened to the $z$ and the $x$ axis,
respectively. (f) A schematic energy level in the AFM/AFO configuration with
the Jahn-Teller-like distortion. The $U-J+3E_{ph}$ excitation appears
instead of the $U-J$ excitation.}
\label{Fig:3}
\end{figure}
\begin{figure}[tbp]
\caption{$T$-dependent spectral weights and $\omega _{TO}$. $S_\alpha (T)$
are subtracted by the value at 100 K to give $\Delta S_\alpha (T)$ $%
(=S_\alpha (T)-S_\alpha (T$=100K$))$, and $S_\beta (T)$ are subtracted by
the value at 220 K to give $\Delta S_\beta (T)$ ($=S_\beta (T)-S_\beta (T$%
=220 K$))$.}
\label{Fig:4}
\end{figure}


\begin{references}
\bibitem{Comment exp1}  For metallic samples, $\sigma (\omega )$ between 4.2
K and 300 K were obtained, but showed little $T$-dependence. And, the Ca$_2$%
RuO$_4$ sample was too small for us to take its $T$-dependent spectra.
\bibitem{Comment1}  The energy costs given in Fig. 3 were evaluated in the
atomic picture. If the full multiplicity is taken into account, the $d^3$
final state in the FM/AFO configuration corresponds to the term $^4$A$_2$,
an eigenstate of the $t_{2g}^3$ configuration. [See Ref. \cite{Degenerate HM}%
.] Then, the excitation energy for the FM/AFO configuration should be $U-4J$.
\bibitem{Comment Fit}  The fittings with Gaussian functions also give
similar behaviors of $S_\alpha $ and $S_\beta $, although their absolute
values are different by about 50 \%.
\end{references}
\end{document}